# Quantum Uncertainty Considerations for Gravitational Lens Interferometry

Laurance R. Doyle[1] and David P. Carico[2]

[1]SETI Institute, 515 N. Whisman Ave., Mountain View, California 94043, ldoyle@seti.org

[2]Department of Earth and Space Sciences, College of the Siskiyous, 800 College Avenue, Weed, California 96094, carico@siskiyous.edu

**Abstract**
The measurement of the gravitational lens delay time between light paths has relied, to date, on the source having sufficient variability to allow photometric variations from each path to be compared. However, the delay times of many gravitational lenses cannot be measured because the intrinsic source amplitude variations are too small to be detectable. At the fundamental quantum mechanical level, such photometric "time stamps" allow which-path knowledge, removing the ability to obtain an interference pattern. However, if the two paths can be made equal (zero time delay) then interference can occur. We describe an interferometric approach to measuring gravitational lens delay times using a "quantum-eraser/restorer" approach, whereby the time travel along the two paths may be rendered measurably equal. Energy and time being non-commuting observables, constraints on the photon energy in the energy-time uncertainty principle—via adjustments of the width of the radio bandpass —dictate the uncertainty of the time delay and therefore whether the "path taken" along one or the other gravitational lens geodesic is "knowable." If one starts with interference, for example, which-path information returns when the bandpass is broadened (constraints on the energy are relaxed) to the point where the uncertainty principle allows a knowledge of the arrival time to better than the gravitational lens delay time itself, at which point the interference will disappear. We discuss the near-term feasibility of such measurements in light of current narrow-band radio detectors and known short time-delay gravitational lenses.

## 1. Introduction

Time delays for a number of gravitational lenses have been measured by correlating flux variability between a pair, or group of gravitational lensed images (e.g., Haarsma *et al*. 1999; Biggs *et al*. 1999, Kundic' et al. 1997, Saha et. al. 2006). This approach requires the presence of moderately short-term detectable variability in the source galaxy or quasar in order for a detectable time-series correlation to be made—a situation which is not always met. However, an interferometric approach, and application of the quantum uncertainty principle, may be able to mitigate this situation.

John Wheeler (1978) first suggested a *Gedanken* experiment in which gravitational lens interferometry could be used to illustrate a delayed-choice paradox. In this *Gedanken* experiment two separate light paths (A and B) around a gravitational lens are allowed to interfere. A "choice" is made by the photons to take either path A or path B if either one of these paths has a measurement device along its path before combining. Alternatively, a given photon might be said to take both paths (Dirac 1958) in the case of interference being measured at their intersection point (to put the phenomenon in a somewhat classical context). Thus, considering just two paths, the photon path choice (A, B, or both paths A and B) is made long after the photon is supposed to have left its source. In Wheeler's original delayed-choice experiment the issue of coherence, required for interference, was addressed by imagining an immensely long fiber optics cable to assure that the light paths along routes A and B were equal. As we discuss below, such an impractically long fiber optics cable may be replaced with an extremely narrow band radio wavelength filter, thereby using the uncertainty principle to "erase" any path-length differences, and remove "which-path" knowledge of the photon's route to the detector by making the time delay between them unmeasureable.

On the assumption that coherence could be attained, Peterson and Falk (1991) suggested using a gravitational lens interferometer for detecting previously unresolved lensed images. Other authors (including Schneider and Schmid-Burgk 1985; Labeyrie 1993, 1994; Gould and Gaudi 1997) have discussed gravitational lens interferometry in detail as well, indicating, however, that such interferometry might be possible only at femto-second to microsecond delay times and only for small radii objects—such as a pulsar in a gravitationally lensed configuration (e.g., Schneider et al.1999). Thus the issue of coherence is an important point that needs to be addressed.

**2. Coherence Considerations: Temporal Coherence**

Given a simple double-slit arrangement, frequency coherence is obtained when one has a sufficiently small bandpass such that the electromagnetic waves being detected from the two paths have about the same wavelength, insuring that the interference patterns being produced all have the same fringe spacings and so are detectable. If the waves are not of the same wavelengths (not coherent in frequency) then interference patterns of different spacings will occur for these different wave lengths, and each of the many differently-spaced interference patterns will overlap, thereby reducing fringe visibility.

Assuming that frequency coherence is satisfied, temporal coherence is achieved when the path length difference between the two paths is less than the coherence length of the source. However, if two paths from a given source arrive at significantly different angles from each other, then even if frequency and temporal coherence are satisfied, the location on the detector of these interference patterns will differ, again reducing the fringe visibility. In this case spatial coherence has not been satisfied. Assuming frequency coherence is satisfied, we note that spacial incoherence is similar to temporal incoherence in that the two paths under consideration are out of phase.  However, spacial incoherence is due to the source being off-axis (including the effects of being an extended source) rather than any intrisic phase "sputtering" at the source itself.

From the quantum mechanical viewpoint, P.A.M. Dirac argued, from conservation of energy considerations, that "Each photon then only interferes with itself. Interference between two different photons never occurs" (Dirac 1958, p. 9). In this regard, a given photon probability distribution can be thought of as "taking both paths" and thus can result in a particular interference pattern after many such photons have been allowed to "build up" on the detector. Since the interference fringe spacing depends on the energy of the photons, fringe visibility demands that the photons all have the same, or nearly the same energy, i.e. frequency coherence. Likewise, another photon probability distribution (wave), coming from a significantly different angle, will make the same interference pattern, but offset from the first. Hence, as before, spatial coherence must be satisfied for the interference pattern to become apparent. Thus both frequency and spatial coherence in both the classical and quantum cases can be understood on the same basis. Interference will be produced in each case but detection of the interference patterns requires a narrow bandpass and approximately equal paths.

In contrast, temporal incoherence (a "sputtering" source) has no classical counterpart at the single photon domain, but can be understood only as the result of the energy-time uncertainty relation. In this regard, if the difference in path length corresponds to a difference in light travel time that is less than the minimum time that can be measured according to the uncertainty principle, then the path length difference is unmeasureable (i.e., unknowable) and so, to the photon, does not exist. If no path length difference is, even in theory, distinguishable, then the paths are measureably equal and interference will occur at the detector.

Thus, while the effects of frequency incoherence can be mitigated by narrowing the bandpass in both the classical as well as the quantum or single photon cases, it is only from the quantum mechanical point of view that narrowing the frequency bandpass can also produce temporal coherence, and it will do this by way of the uncertainty principle. Since one cannot measure the difference in path lengths (i.e., have knowledge of which path a photon took) more precisely than the uncertainty principle allows, this uncertainty may be applied to manipulate the conditions needed for interference by imposing ignorance on which-path information. As we shall see below, the point at which coherence occurs should allow the direct measurement of the delay time along the two gravitational lens light paths irrespective of source variability (i.e., without time "tags" from the source).

**3. Classical Gravitational Lens Interferometry**

The delay time between two gravitationally-lensed point-source light paths can be most generally formulated as:

$$\Delta \tau = (1+z) \left[ \frac{D_s D_l}{2 D_{sl}} (\theta_A - \theta_B)^2 - \frac{\Phi(\theta)}{c^3} \right] \qquad (1),$$

where the affine path distance from the observer to the lensed source is $D_s$, from the observer to the lens is $D_l$, and from the source to the lensing object is $D_{sl}$ (after Burke and Graham-Smith 1998; see also Schneider et al. 1999). The angle between the observer's line-of-sight to the center-of-mass of the lens and any given lens-produced image is $\theta$ (the angle to either image A or image B for the two-image system of Equation 1), $z$ is the redshift of the lensing galaxy and $c$ is the speed of light. The first generalized term within the brackets represents the component of the time delay due to geometric considerations of path-length, while the second term, $\Phi(\theta)/c^3$ (typically of the same order of magnitude as the first term) represents the relativistic effects of the gravitational potential well of the lens, contributing to the time delay via time dilation. (For now we shall assume that the uncertainty principle will "see" the time dilation component as well as the geometric path length component in our considerations below— i.e. that the uncertainty principle is effected by general relativistic considerations of time dilation, but we discuss this again below.)

For a point mass deflection:

$$\theta = \frac{\beta}{2} \pm \left[\frac{\beta^2}{4} + \left(\frac{D_{sl}}{D_s D_l}\right)\left(\frac{4GM}{c^2}\right)\right]^{\frac{1}{2}} \qquad (2),$$

where $M$ is the lensing galaxy's mass, $\beta$ is the angle from the source to a massless lens, (i.e., the angle to the lensing galaxy's position from the image(s) source in the absence of any gravitational deflection), and $G$ is the universal gravitational constant. For the case where the lensing mass is aligned with the source along the observer's line of sight ($\beta = 0$), an Einstein ring will result, with an angular radius (from Equation 2) of:

$$\theta_E = \sqrt{\left(\frac{4GM}{c^2}\right)\left(\frac{D_{sl}}{D_l D_s}\right)} \qquad (3).$$

As mentioned, classically interference has been thought of as being produced by the superposition of coherently produced waves (time coherence) of nearly the same wavelength (frequency coherence) and from a source that subtends a very small angular distance across the plane of the sky (spatial coherence). In the case of a standard Young's double-slit arrangement, (i.e., a gravitational lens with two superimposed images A and B along position vectors $\vec{r}_A$, $\vec{r}_B$ and travel times $t_A$, $t_B$ from the source), the average intensity distribution, $\langle I(\vec{r},t) \rangle$, for a stationary field (i.e., $t - t_A = t - t_B = 0$), in terms of the normalized first-order correlation function, $g^{(1)}(\vec{r}_A, \vec{r}_B, \Delta\tau)$, will be:

$$\langle I(\vec{r},t) \rangle = \langle I(\vec{r}_A) \rangle + \langle I(\vec{r}_B) \rangle + 2[\langle I(\vec{r}_A) I(\vec{r}_B) \rangle]^{1/2} \times |g^{(1)}(\vec{r}_A, \vec{r}_B; \Delta\tau)| \cos[\alpha(\vec{r}_A, \vec{r}_B; \Delta\tau) - \nu_0 \Delta\tau] \quad ,$$

where $\Delta\tau = |t_A - t_B|$ is the total gravitational delay time (as given in Equation 1), $\nu_0$ is the quasimonochromatic radio field frequency being observed, and

$\alpha(\vec{r}_1, \vec{r}_2; \Delta\tau) = \arg[g^{(1)}(\vec{r}_1, \vec{r}_2; \Delta\tau)] + \nu_0 \Delta\tau$ (e.g., Mandel and Wolf 1995, Scully and Zubairy 2001). For very narrow-band radio flux measurements, $\langle I(\vec{r}_A) \rangle$, $\langle I(\vec{r}_B) \rangle$, $|g^{(1)}(\vec{r}_A, \vec{r}_B; \Delta\tau)|$, and $\alpha(r_A, r_B; \Delta\tau)$, should vary slowly with respect to position on the detector, while the cosine term should vary rapidly due to the $\nu_0 \Delta\tau$ term, leading to sinusoidal variations of intensity at the radio telescope array (e.g., Scully and Zubairy 2001).

For the simple Young's double-slit experimental setup, the sharpness of the interference pattern is defined by the fringe visibility:

$$V = \frac{\langle I(\vec{r}) \rangle_{max} - \langle I(\vec{r}) \rangle_{min}}{\langle I(\vec{r}) \rangle_{max} + \langle I(\vec{r}) \rangle_{min}} \quad ,$$

where "max" and "min" represent the maximum and minimum average intensities at the detector. To a good approximation, $\cos[\alpha(\vec{r}_A, \vec{r}_B; \Delta\tau) - \nu_0 \Delta\tau] = \pm 1$, for the maxima and minima, respectively. Thus the fringe visibility in terms of the complex degree of coherence, $g^{(1)}(\vec{r}_A, \vec{r}_B; \Delta\tau)$, can be written as:

$$V = \frac{2[\langle I(\vec{r}_A)\rangle \langle I(\vec{r}_B)\rangle]^{1/2}}{\langle I(\vec{r}_A)\rangle + \langle I(\vec{r}_B)\rangle} |g^{(1)}(\vec{r}_A, \vec{r}_B; \Delta\tau)| \qquad (4).$$

For a Doppler-broadened thermal light source (i.e., stellar-like sources) the first-order correlation function can be described by:

$$|g^{(1)}(r_A, r_B; \Delta\tau)| = \exp\left[-(\Delta\tau)^2 / 2(\Delta\tau_c)^2\right] \qquad (5),$$

where, for now, $\Delta\tau_c$ is a constant. As the path difference, $c\Delta\tau$, becomes much larger than a critical path length, $c\Delta\tau_c$, Equation 5 then goes to zero and the classical interference fringes disappear.

For the purposes of outlining the constraints imposed only by the uncertainty principle, we now make some simplifications. In the radio regime, effects of interstellar scintillation, caused by the intergalactic/interstellar medium, the solar wind, and the terrestrial ionosphere, will affect the fringe visibility by superimposing the effects of a corrugated wave-front on the detector (i.e. another faint fringe pattern), as well as shifting phases and broadening frequencies. These effects typically may increase natural spectral line widths in the radio regime to greater than about $10^{-2}$ Hz (e.g., Burke and Graham-Smith 1998, Walker 1989, and references therein). We also recognize that the brightness distributions of the gravitational lens images are dependent on the caustic(s) through which the light passes and that this can change (albeit slowly with respect to the speed of light). In this present paper (again for simplicity) we shall assume that these external limiting factors to detection of interference can be practically mitigated by observational techniques. (For example, that the changing index of refraction of interstellar plasmas would produce an interference pattern at sufficiently different frequencies as to be distinguishable from the interference fringes that we are producing between source images A and B, etc.).

Interference will therefore be, at its simplest, a function of the coherence conditions and the interferometer path-length difference. The coherence length of a source is dictated by the correlation function (as given in Equation 5), so that measurement of the interference fringes of a thermal source would generally be limited in a gravitational lens interferometer to an angular diameter $\delta$ of about:

$$\delta \leq 10^{-17} \frac{H_0}{\nu} \left( \frac{D_l D_s}{M D_{ls}} \right)^{\frac{1}{2}} \qquad (6),$$

where $H_0$ is the Hubble constant (e.g., Schneider et al. 1999, Gwinn et. al. 2000). For a lensing source even as small as one solar-mass at a radio wavelength on the order of a few centimeters we see that the source would have to be smaller than about $10^{10}$ cm in order to obtain interference fringes in a Young's-type double-slit experimental setup. Thus a double-slit experiment in gravitational lens interferometry would be limited, for example, to eclipsing binary pulsars emitting coherent radiation or perhaps possibly a symmetric Einstein ring at very long-baseline resolution. However, as discussed, this is not a limitation to detecting interference in a Mach-Zehnder interferometer where spatial coherence is already assured and temporal coherence is governed by the uncertainty principle.

A concern arises with regard to the detection of extended sources in which various interference fringes would be expected to overlap with each other due to spatial incoherence, so that interference could not be detected. However, in a Mach-Zehnder interferometer configuration one gives up spatial information (about the angular distribution of the extended source on the sky) for the simple detection of interference itself, replacing a fringe visibility pattern with a simple photon counter. Therefore we discuss coherence considerations for a Mach-Zehnder interferometer configuration in the next section.

**4. Extended-Source Coherence and the Mach-Zehnder Interferometer**

In a standard laboratory set up for a Mach-Zehnder interferometer light from a source is sent along two distinct paths, A and B, after encountering a beam splitter. With the use of mirrors these two paths are brought to a point of intersection where they encounter a second beam splitter. Two detectors then record the light that either is reflected or passes through the second beam splitter. For the case where the two paths are identically equal, under ideal conditions all of the light emitted by the source will be directed by the second beam splitter to one of the detectors, and none of it to the other detector.

In practice, because even a collimated laser beam must have some finite width, not every pair of paths A and B from an extended source to the second beam splitter will have identically equal lengths. In fact, only one such pair of paths can exist for each point on the source. This pair of paths will produce a bright spot on one detector and nothing (that is, a dark spot) on the other. However, any two neighboring paths, A´ and B´, starting from the same point on the source, will have slightly different lengths and hence will interfere differently. Since light traveling along paths A´ and B´ will arrive at the detectors from slightly different directions—and hence hit the detectors at slightly different places than light from paths A and B—the resulting interference will have no effect on the bright and dark spots produced by paths A and B. Rather, the result of these various pairs of paths all interfering at the detectors is a series of concentric rings, a "bull's eye pattern," on each detector, one with a bright center and the other with a dark center. But this is solely due to the finite aperature size of the collimation.

Since the location of the central bright or dark spot is independent of the frequency of the light, it is not essential that the source have a narrow frequency bandwidth—that is, frequency coherence is not required. All wavelengths emitted by a given point on the source and traveling along the same identical paths will produce the same interference "hits" (or lack of hits) at the two detectors. However, since the spacing of the interference rings is a function of the frequency, a narrow frequency bandwidth can considerably enhance the visibility of the interference effect by essentially eliminating all but one set of rings. In the laboratory this is usually achieved with the use of a fine tuned laser, although a sufficiently narrow filter at the source, or a pair of filters positioned anywhere along the two paths, can also be used.

The size of the source is also of importance in observing interference. Each distinct point on the source will produce an interference ring pattern on each detector that is centered on a different location. Hence, in order for these interference patterns not to overlap—which would reduce or eliminate their visibility—it is necessary that the source be either nearly a point source, or if it is extended, that the light be well collimated, that is, spatially coherent.

Finally, in order for the interference to be stable for a measurable period of time, the two paths must have the same or nearly the same length. That is, temporal coherence must be established and maintained. If a path extension or delay line is inserted into one of the paths (e.g., fiber optics cable for optical experiments), interference will begin to disappear when the path length difference approaches the coherence length of the source, and will disappear completely as this length is exceeded.

These considerations can be carried over to a gravitational lens. To begin let's assume that the source is a point source, thus insuring spatial coherence. The first beam splitter is provided by the gravitational lens, which defines at least two paths from the source to the telescope. The probability wave of each photon will "split" upon encountering the lens, since in the absence of "which path" information the photons are required to "travel" both paths. The gravitational lens "beam splitter" will have an effective refractive index of (Schneider et al. 1999, p. 123):

$$n = 1 - \frac{2U}{c^2} + \frac{4}{c^3} \vec{V} \cdot \vec{e} \qquad (7),$$

where $U$ is the Newtonian potential of the lensing mass distribution, $\vec{V}$ is the gravitational vector potential, and $\vec{e}$ is the unit tangent vector a ray. Thus the phase of the wave will be expected to change as a result of the gravitational lens. However, this should not obstruct the detection of interference (as we consider below).

The final elements of the interferometer are provided by the observer. Each beam is first passed through a narrow band filter to establish frequency coherence. The two beams are then allowed to intersect, a beam splitter (the second beam splitter in a standard Mach–Zehnder interferometer) is placed at the point of intersection, and the light emerging from the beam splitter is directed towards two detectors .

The principle difficulty in using a gravitational lens as a Mach-Zehnder interferometer is that, unlike the situation in the laboratory, the two paths will in general have very different path lengths. Typical measured path length differences range over many days (although it is important to note that in the few known Einstein rings, nearly identically equal paths may be possible to identify; see discussion below). In his original thought experiment, John Wheeler acknowledged this difficulty by including a delay line in one path (see Wheeler and Zurek, 1983). Although this kind of straightforward solution is clearly impractical, we argue that it is nevertheless possible to render the two paths measurably equal, thus insuring temporal coherence, by utilizing the uncertainty principle. As we discuss in detail below, by narrowing the frequency bandwidth sufficiently, it should be possible, by way of the well known energy-time uncertainty relation, to render the path length difference between two gravitationally lensed paths unknowable, and hence unmeasurable. If this path length difference is unmeasurable, then it can have no measurable effect on any observations; which is to say that it effectively does not exist. The two paths are identically equal then as far as any measurement is concerned, and temporal coherence is thereby established.

The time-energy uncertainty principle is rarely the limiting factor in obtaining interference, and never a factor in optical interferometry. However, conditions may be set up so that this constraint does take precedence at radio wavelengths, and thus may be used to remove certainty about travel time differences along light paths, thereby allowing interference to take place. We propose a quantum restorer—in the sense that we propose to "erase" an interference pattern once obtained, thus returning the system to the quantum (i.e. knowledge of which-path) state, enabling direct measurement of the gravitational lens delay time. We then discuss aspects of the potential realization of such an experiment using two examples—that of an eclipsing pulsar and that of an Einstein ring. It may be noted in passing that uncertainty introduced into the which-path information does not just allow interference, but *requires* it as, for example, in the well-known case of quantum beats, which cannot be explained classically (e.g., Greenstein and Zajonc 1997, Scully et. al. 1991, Scully and Zubairy 2001).

**5. Knowability and Detection of Interference**

From Equation 5 we see that, in addition to the wave coherence time, the correlation coefficient also depends on the magnitude of the delay time itself since the uncertainty principle must be satisfied. This constraint usually applies only to the radio region of the spectrum as this region allows the narrowest relative bandpasses. Since:

$$\Delta\lambda = \frac{-c\Delta\nu}{\nu^2}, \qquad \text{for small } \Delta\nu, \qquad (8)$$

if one wants to use the uncertainty principle to manipulate a delay time of, for example, 1/10th second (i.e., $\Delta v = 10$ Hertz), then at visible wavelengths (0.55 microns) the filter bandpass would have to be adjustable to within about $\Delta \lambda = 1.35 \times 10^{-14} \mu m$, or substantially less than the width of an atomic nucleus. On the other hand, at moderate radio wavelengths of about $\lambda = 3$ cm, the band-width of a 10 Hz filter would be about 0.3 mm. Thus, in the radio region of the spectrum, the existence of interference may be affected by conditions defined by the complimentarity of energy and time, which results in a well-known fundamental quantum limitation in radio astronomy, the minimum time required by the uncertainly principle being about:

$$\Delta t_{UP} = \frac{1}{2\pi \Delta v} \qquad (9).$$

The quantum measurement description of an interference phenomenon makes the distinction between that which is *unknowable*, and that which is merely *unknown* (e.g., Bell 1994). Information which violates the uncertainty principle is unknowable; information which does not violate the uncertainty principle is knowable, although—by choice of experimental conditions—it may be unknown to the observer (e.g., Kim et al. 1999). In order for interference to occur when two light paths from the same source are superimposed, the information as to which path each photon has 'traveled' must therefore be rendered unknowable by the experimental setup. The state of photons for which the path information is unknowable is characterized by the superposition state,

$$|\psi\rangle = \alpha|A\rangle + \beta|B\rangle \qquad (10)$$

where $|A\rangle$ and $|B\rangle$ represent the individual states for photons that have 'traveled' along paths A and B, respectively, and $|\alpha|^2$ and $|\beta|^2$ are the corresponding probabilities associated with each path—which for a two-image gravitational lens are proportional to the relative intensities, $\langle I(\vec{r}_A)\rangle$, $\langle I(\vec{r}_B)\rangle$, of the two images. Hence another way of stating the condition for interference to occur between the two beams is that the photons arriving at the detector must be in the superposition state $|\psi\rangle$ given by Equation 10.

In the standard laboratory double-slit experiment this superposition (i.e., 'unknowability') of the path taken by each photon is achieved by insuring that the two light paths are spatially and temporally indistinguishable to the observer. In an extra-galactic gravitational lens, the light paths are generally of very different lengths, differing by at least many light-days. The two paths are therefore distinguishable, making the path information for each photon knowable and preventing interference from occurring.

Any point source interference pattern will disappear abruptly as illustrated in various laboratory delayed-choice experiments when such 'which-path' information becomes knowable (e.g., Hellmuth et al. 1987, Kim et. al. 1999, Peterson 2002). Scully et al. (1991) have also shown that this effect is independent of any direct interaction with the photon (such as momentum transfer by short-wavelength photons, which was the original mechanism proposed by Heisenberg in the first uncertainty principle paper for the limits on measurement certainty). This uncertainty principle approach will also apply to extended sources where one can expect interference as the additional delay time across the finite source is also "erased" by the uncertainty principle. Let us now examine, in a bit more detail, considerations imposed only by the uncertainty principle upon interference in terms of simple inequalities that determine when interference can be observed (i.e., when which-path information is knowable or unknowable).

## 6. Uncertainty Principle Constraints on Interference

Under consideration are the three differential time quantities: $\Delta t_{EXPO}$, which is the exposure time for a given observation of the gravitational lens superimposed images, $\Delta t_{UP}$, which is the uncertainty principle time as defined in Equation 9 above, and $\Delta \tau$, which is the gravitational lens delay time, as defined in Equation 1 above. In the cases below, for a point source, the delay time is the time-travel difference between the two geodesic paths (both the geometric and relativistic time-dilation components are taken together for now). In the case of a more extended source, the total delay time can be assumed to be the point source delay time plus the travel time differences across the extended source which, as mentioned, may also be "erased" by an additional narrowing of the filter bandpass.

Upon superimposing two gravitationally lensed light paths (we assume a Mach-Zehnder interferometer configuration, as discussed previously), consideration of various orderings of these three temporal parameters leads to the following possibilities for or against interference.

Case 1:

$$\Delta\tau > \Delta t_{expo} > \Delta t_{UP} \qquad (11)$$

This is the usual differential time relationship for the observation of a gravitational lens. Since $\Delta\tau > \Delta t_{UP}$, information as to which path each photon has 'traveled' is, in principle, knowable. That is, the arrival time of the photon(s) can be determined to an accuracy that is less than the delay time between the two paths. The paths are therefore distinguishable, and the photons arriving at the detector will consist of a mixture of path possibilities $|A\rangle$ and $|B\rangle$. Thus, although the beams of the sources A and B cross paths at a detector, interference cannot occur because the wave functions $|A\rangle$ and $|B\rangle$ themselves are not in a superposition state. Because $\Delta t_{expo} > \Delta t_{UP}$ this difference will be measurable, and since $\Delta t_{expo} < \Delta\tau$ the gravitational delay time can be directly measured if there is sufficient source intensity variability to be detected via a correspondence between the intensities of the two light-curves.

Case 2:

$$\Delta t_{expo} > \Delta\tau > \Delta t_{UP} \qquad (12).$$

Since $\Delta t_{EXPO}$ is now greater than the delay time of the gravitational lens, this insures that the information about which path the photons may have taken will be unknown to the observer, but only by the observer's choice of exposure time, and not intrinsically by the experimental set-up itself. However, since $\Delta\tau$ is still greater than $\Delta t_{UP}$, the which-path information is not *unknowable* in principle. Hence, again, the photons arriving at the detector will consist of a mixture, rather than a superposition, of the states $|A\rangle$ and $|B\rangle$, and interference will not be produced.

Cases 3:

$$\Delta\tau > \Delta t_{UP} > \Delta t_{expo} \qquad (13a)$$

$$\Delta t_{UP} > \Delta \tau > \Delta t_{\text{exp}o} \tag{13b}.$$

In both of these cases, the exposure time is less than the minimum uncertainty allowed by the uncertainty principle. This can occur in a situation, for example, in optical astronomy, in which the detection of the photons comes *after* the light has already passed through a filtering device. In such cases, the filter may constitute a measurement constraint on the energy of each photon, and the detector provides a subsequent measurement of the time at which each photon had that corresponding energy (given the constant speed of light) if it is close enough to the filter. Since energy and time are noncommuting observables, the measurement of the time of detection must therefore limit the observer's knowledge of each photon's frequency to a value:

$$\Delta \nu > 1/\Delta t_{\text{exp}o} \tag{14},$$

at the time from passing through the filter to registering on the detector (Greenstein and Zajonc 1997, p.60). In other words, the photon's energy distribution can be broadened by the very short exposure time itself, allowing photons of a wider energy dispersion to be detected which, classically, should not have been able to pass through the filter.

The net result is that the observer's knowledge of the *time* at which a given photon had an energy within the range $\Delta E = h\Delta \nu$ is limited by $\Delta t_{\text{exp}o}$ rather than $\Delta t_{UP}$. Hence, since $\Delta \tau > \Delta t_{\text{exp}o}$ in both cases, interference will not occur for the same reasons as given in Case 1. We note that for Inequality 13b, because , $\Delta t_{UP} > \Delta \tau$, interference might at first be expected to be detectable. However, since the exposure time is smaller than the uncertainty principle time, the narrowness of the bandpass should not be dictated by the uncertainty principle time itself, as noted, but rather limited by the shortness of the exposure time (i.e., a very rapid exposure time will broaden the bandpass for the same reason it does in Inequality 13a). Thus $\Delta \nu$ is not as narrow as the uncertainty principle would dictate, but only as narrow as exposure time allows it to be, and this will essentially make the uncertainty principle time smaller than the delay time again, regardless of the classical bandpass of the filter.

Cases 4:

$$\Delta t_{\text{exp}o} > \Delta t_{UP} > \Delta \tau \tag{15a}$$

$$\Delta t_{UP} > \Delta t_{\text{exp}o} > \Delta \tau \tag{15b}$$

In the first inequality, since $\Delta t_{UP} > \Delta \tau$, the path information for each photon is fundamentally unknowable, so that interference can take place. Also, since $\Delta t_{\exp o} > \Delta t_{UP}$ photons within the energy range $h\Delta v$ can be detected and constitute the interference pattern. In the second inequality, since $\Delta t_{UP} > \Delta t_{\exp o}$, the temporal uncertainty is provided by $\Delta t_{\exp o}$, but again since the limiting uncertainty is greater than the delay time, path information is unknowable, and interference can occur. However, again, a very short exposure time can broaden the energy distribution of the photons reducing the detectability of interference somewhat by compromising frequency coherence.

These inequalities specify the necessary constraints placed by the uncertainty principle, for the detection of interference between superimposed beams from paths A and B (here, from a gravitationally lensed source). The essential uncertainty principle requirement is that the limiting temporal uncertainty, $\Delta t_{UP}$, be greater than the delay time of the gravitational lens (Inequalities 15), with interference expected to be most clearly detectable when the conditions of Inequality 15a are ideally met.

**7. A Quantum "Restorer" Approach**

The usual method for determining delay times between gravitationally lensed sources (generally with the goal of determining a more precise value for the Hubble constant , the lensing object's density distribution, etc.) is to attempt to correlate the optical, infrared, or radio brightness variability of each source image with the other source images under the conditions of Equation 4. Although this has been successfully performed for some gravitational lens systems, in quite a few others the light source has not been sufficiently variable to allow such an intensity correlation of the time series measurements to be performed.

When the conditions of Inequality 15a are achievable, however, a quantum restorer (i.e. interference eraser) can thus be applied as follows. The gravitationally lensed beams are superimposed, beginning with a radio-wavelength bandpass that is sufficiently broad to insure that no interference can occur (i.e., "which-path" information is knowable). The bandpass, $\Delta v$, is then stepwise narrowed until the interference pattern begins to appear at a critical frequency bandpass, $\Delta v_c$. This will occur at the point where increased certainty in the knowledge of the photon energy, $\Delta E$, forces a sufficiently large uncertainty in the arrival time of the radio photons so as to preclude any knowledge of which path they have taken. In other words, the interference will disappear when $\Delta t_{UP}$ becomes greater than $\Delta \tau$. The delay time is then measurable as:

$$\Delta \tau \approx 1/2\pi \Delta \nu_c \qquad (16).$$

If the onset of interference is not detected after the bandpass has been narrowed as much as possible, then the minimum bandpass can be used to establish a lower limit on the actual delay time. A semi-classical way of looking at this process is that, as the bandpass is narrowed and the minimum measurable time interval, $\Delta t_{UP}$, is "stretched" the probability distributions (along paths A and B) are "stretched" as well, and hence they overlap more and more. The region of overlap represents the interval over which the photon's path is unknowable, which is the fundamental condition for interference. The more the probability distributions overlap, the more photons will manifest an interference effect.

As intimated in Section 5, this methodology should also work to "erase" the additional time delay from separate regions across extended sources— limited only by the narrowness of the bandpass achievable. Since each point on an extended source will correspond to a slightly different delay time, self-interference of individual photons from different points will be erased at slightly different frequency bandpasses, with the distances points with the smallest delay time between them being "erased" first. Hence, a measure of the range of frequencies over which interference is detected could enable an estimate of the extended size of the source itself. Present technology may allow the size of quasars, for example, to be measured in this way, although extended radio flux from whole galaxies at this point would be problematical, given the present unavailability of such extremely narrow radio bandpasses as would be required to apply this technique.

## 8. Gravitational Lenses with Small Delay Times

The narrowest radio bandpasses in use today are on the order of $10^{-3} - 10^{-4}$ Hz (used, for example, for the search for extraterrestrial intelligence; Tarter 2001). Hence, gravitational lenses to be measured as a test of this approach must presently be limited to relatively small delay times of 3 hours. At present this method may be testable on two kinds of sources: eclipsing binary pulsars and very symmetrical Einstein rings.

The discovery of a double pulsar PSR 0737-3039 (Burgay et al. 2003, Lyne et al. 2004, Kalogera et al. 2004) which is almost edge-on (3° orbital inclination to our line-of-sight at the time of its discovery) may allow a test of this method. These two pulsars nearly eclipse each other, and the orbital nodes are precessing toward an even smaller orbital inclination line-of-sight angle at a very rapid rate (Lyne et al. 2004). A time delay (the Shapiro-delay, due to relativistic time dilation) of about $10^{-4}$ second, along with the geometric path length, has already been detected along one path (from PSR 0737-3039A), requiring a relatively small but reasonable radio bandpass of 10 kHz to satisfy Inequality 15a. Coherent pulsar flux would be of substantial assistance in detecting interference in this case.

Perhaps more difficult would be the effort to detect interference in closely aligned gravitational lens quasar/galaxy configurations such as the symmetric Einstein ring B1938+666. Sufficient ground-based resolution with large radio telescope arrays may nevertheless allow delay times of about 100 seconds to be measured by interfering opposing sides of such precisely aligned gravitational lens systems, thus requiring a $\Delta \nu \leq 5 \times 10^{-3}$ Hz radio filter. Thus a large radio interferometer might just resolve one side of the Einstein ring allowing the flux to interfere with the symmetrically opposite region on the other side of the ring (interference would occur when this matching region was detected). Of additional concern in such an experiment, however, is the size of the telescope required, as such narrow bandpass observations would require long exposure times. For two such Einstein ring sub-regions (i.e. areas on either side of the annulus) made to interfere, the integration time, $\Delta t_{\exp o}$, required for a one-sigma level detection can be estimated by:

$$\Delta t_{\exp o} = \frac{2k^2 T_{n1} T_{n2}}{A_1 A_2 \Delta \nu (\Delta S)^2} \quad (17),$$

where $T_{n1}$ and $T_{n2}$ are the assumed antennae noise temperatures, $A_1$ and $A_2$ are the areas of the antennae, $\Delta v$ is the bandpass, $\Delta S$ is the source flux, and $k$ is Boltzmann's constant (after Burke and Graham-Smith 1998). Approximating the square kilometer array (SKA) as two very large antennae of equal areas (ignoring the filling factor, for now), using noise temperatures of 30 K each, a detectable source flux of about 50 milliJanskys (e.g., Einstein ring B1938+666, components C1 and C2 at 5 GHz), and a band pass of $10^{-4}$ Hertz, the integration time required would be on the order of a day. Considerations (such as the filling factor) might increase this to more than two weeks (i.e. circumpolar objects would preferentially be observed). As noted earlier, interstellar scintillation and other effects would have to be considered as well in a practical observing program to test this methodology. To extend this method to longer gravitational lens delay times will require advances in narrow-band radio detector technology as well as refinements of observing techniques to overcome noise sources.

## 9. Discussion and Conclusions

A future experiment utilizing the method outlined herein could constitute a realization of the original delayed choice *gedanken* experiment proposed by John Wheeler (1978). Furthermore, the cosmic scale of such an experiment might enable empirical exploration of any distance dependence on the rate at which a photon "changes" from the superposition state to the mixed state (see e.g., Greenstein and Zajonc, 1997; Kim et al. 1999, Callender and Huggett 2001). It would be of interest if such a minimum limit on the time required to produce or eliminate an interference pattern using this method were found to be dependent upon the distance to the gravitational lenses themselves, irrespective of the gravitational lens delay times measured. A lower limit on this "rate" (of about 10,000 times the speed of light) has recently been placed on this "spooky action at a distance" by Salart et al. (2008).

Finally, the assumption that the uncertainty principle is subject to the general relativistic gravitational well time dilation—and not just the geometric-length portion of the delay time—might also be tested. As noted, a significant portion of the delay time of gravitational lenses can be due to the time spent (so to speak) by the photon in a large gravitational potential well. This time dilation effect is due solely to a relativistic gravitational potential well, while the uncertainty principle has perhaps been thought to relate—via the non-commutability of energy with time—to geometric considerations alone. By applying this technique to known gravitational lenses, it would be interesting to note whether the quantum eraser outlined here also includes in its erasure process the time delay due to the gravitational potential well also. Such an experiment might supply an interesting experimental connection between quantum physics and general relativity.

In conclusion, we hope that the manipulation of the uncertainty principle to erase and restore path length knowability in interferometric systems can eventually become a useful methodology for measuring delay times in non-variable gravitational lens sources, and also find application to other astronomical-scale experiments in fundamental physics.

## 10. Acknowledgements